\documentclass[%
 reprint,
 amsmath,amssymb,
 aps,
prl,
floatfix,
]{revtex4-2}

\usepackage{graphicx}% Include figure files
\usepackage{dcolumn}% Align table columns on decimal point
\usepackage{bm}% bold math
\usepackage{float}
\usepackage{xcolor}
\usepackage{soul}
\newcommand{\iamd}{$Ia\overline{3}d$}

\begin{document}

\preprint{APS/123-QED}

%\title{Unexpectedly complex phase behavior of a simple model system provides deeper understanding of liquid polymorphism}
\title{Insight into liquid polymorphism from the complex phase behaviour of a simple model}
%\title{Surprises of the liquid-liquid phase transitions}% Force line breaks with \\

\author{Albert P. Bart\'ok}
\affiliation{Department of Physics and Warwick Centre for Predictive Modelling, School of Engineering, University of Warwick, Coventry, CV4 7AL, UK}
\author{Gy\"orgy Hantal}%
\affiliation{Institute of Physics and Materials Science, University of Natural Resources and Life Sciences, Peter-Jordan-Strasse 82, 1190 Vienna, Austria}
%\affiliation{%

\author{Livia B. P\'artay}
\email{Livia.Bartok-Partay@warwick.ac.uk}
\affiliation{Department of Chemistry, University of Warwick, Coventry, CV4 7AL, UK}

\date{\today}% It is always \today, today,

\begin{abstract}
We systematically explored the phase behavior of the hard-core two-scale ramp model suggested by Jagla[E. A. Jagla, Phys. Rev. E 63, 061501 (2001)] using a combination of the nested sampling and free energy methods. 
The sampling revealed that the phase diagram of the Jagla potential is significantly richer than previously anticipated, and we identified a family of new crystalline structures, which is stable over vast regions in the phase diagram.
We showed that the new melting line is located at considerably higher temperature than the boundary between the low- and high-density liquid phases, which was previously suggested to lie in a thermodynamically stable region.
The newly identified crystalline phases show unexpectedly complex structural features, some of which are shared with the high-pressure ice VI phase.

%We systematically explored the potential energy landscape of the hard-core two-scale ramp model suggested by Jagla[E. A. Jagla, Phys. Rev. E 63, 061501 (2001)] with the nested sampling method. 
%The sampling revealed that the phase diagram of the Jagla potential is significantly richer than previously anticipated, and we identified a new stable crystal structure, having a higher melting temperature than that of the previously considered hcp phase.
%To overcome technical challenges posed by the substantial density difference between liquid and the new crystalline structures, we also performed two-phase coexistence Gibbs-ensemble Monte Carlo simulations, and thermodynamic integration, to calculate the melting line.
%We showed that the new melting line lies at considerably higher temperature than the LDL and HDL phase boundary, suggesting that although the Jagla has been considered as the only isotropic model having a liquid-liquid critical point in the stable region of the phase diagram, this is not the case.  

\end{abstract}

\maketitle

%\tableofcontents
%\section{Introduction}

Polyamorphism, the occurrence of a material in more than one non-crystalline form, such as amorphous solids or distinctly different liquid phases, keeps fascinating materials scientists.
While such phases do not exhibit long range order, their local atomic level structure can be markedly different, resulting in different physical characteristics, such as the density or electronic properties.
The existence of multiple disordered phases, especially that of liquid phases with different densities, and the first order phase transitions between them have been long suspected to have a far reaching influence on materials properties, for example potentially explaining the high pressure melting line maximum observed in certain metals.\cite{Rapport} 

Due to the substantial technical challenges, experimental observation of the peculiar liquid-liquid polyamorphism is rare in one-component systems.
Over the past two decades its existence has been suggested only for a few such materials: phosphorus\cite{phosphorous_first-order_2000} silicon,\cite{silicon_liquid-liquid_2003,poole_phase_1992,sciortino_physics_2003}, nitrogen\cite{nitrogen_LL_2007}, cerium\cite{cerium_first-order_2013} or triphenyl phosphite,\cite{tanaka_LLP} as well as the most widely studied material long suspected to have a liquid-liquid phase transition, water.\cite{mishima_h2o_1998,Mishima2010,Singh2017}

Besides a surprisingly complex phase diagram (with so far 19 identified solid phases \cite{delRosso2016,ice_xix}), water is also known to display a wealth of anomalous properties compared to other simple liquids. 
These include the non-monotonous temperature dependence of the density and the viscosity, as well as the increasing isobaric specific heat and isothermal compressibility when the temperature is decreased in the deep supercooled regime. 
This latter phenomenon can be interpreted as indicative of a potential phase transition in the metastable liquid state upon further cooling.\cite{Angell1973,Speedy1976}
The unconventional behavior of water is understood to stem from the fact that hydrogen bonding, which has a heavily directional character, is a major contributor to the cohesive interaction between molecules.
This has two important consequences for the structure of water: bonding distances are very short and the first neighbour shell comprises, on average, only 4 molecules, in contrast to 12 in case of most other simple atomic or molecular liquids.
At ordinary conditions, this dominance of hydrogen bonds gives rise to a strongly bound open tetrahedral structure of relatively low density, but it was suggested that larger pressures lead to the partial collapse of the hydrogen bonding network, resulting in higher coordination numbers.\cite{debenedetti_review,water_pali}
As a consequence, the equilibrium between these two competing local arrangements can bring about the existence of the two distinct liquid phases: the low- and high-density liquid (LDL and HDL),\cite{Cuthbertson2011,Russo2014} 
explaining numerous aspects of water's anomalous behaviour,\cite{Stanley1980} and consistent with further experimental observations of two amorphous arrested glassy states showing structural similarities with them.\cite{Soper2000,Handle2017}.

However, the suspected liquid-liquid critical point (LLCP), specifically in water, lies so deep in the metastable, supercooled region of the phase diagram, that it is highly challenging to access not just experimentally, but also in computer simulations\cite{debenedetti_review}.
While several studies indicate the existence of a HDL phase of a range of water potential models,\cite{Abascal2010,Ni2016,Biddle2017} so far unequivocal proofs have only been presented in the case of the ST2\cite{tanaka_st2}\cite{Palmer2014} and TIP5P models\cite{LL_tip5p}. 

In order to provide more insight into the properties of such short-lived liquid phases, the attention turned towards designing and studying model potentials, where the LLCP lies above the crystallization boundary, allowing the detailed examination of structural properties and thermodynamic response functions in the stable region of the phase diagram.

It was found that the existence of multiple liquid phases is a generic feature of model tetrahedral network fluids (patchy particles) with an LLCP\cite{Sciortino2010} which can be brought to the stable regime upon increasing the angular flexibility of the model, which lowers the energy cost of the disruption of the network structure.\cite{Smallenburg2015}

Jagla showed in their seminal works that directionality is not a necessary feature for an interatomic potential model to display liquid-liquid polyamorphism. 
%They proposed two potential models, 
They studied the soft-core double step potential\cite{soft-core_LL1_2002,soft-core_LL_2004} and the hard-core two-scale ramp model\cite{Hemmer_jagla,jagla}, which are both monoatomic, isotropic pair potentials.
The latter model, which 
was originally proposed, and its second critical point examined in the case of the one-dimensional fluid, by Hemmer and Stell~\cite{Hemmer_jagla},
became known as the \emph{Jagla model}. 
The Jagla or Hemmer-Stell model has since received increased attention, as it was found that its LLCP appears in the stable liquid regime, and thus has been used as a toy model to understand the polyamorphism of more complex liquids, such as water.\cite{Jagla_DFT,lomba_TI} 
A range of other non-directional soft-core potential models has been since studied,\cite{fomin_review} several of which display anomalous behaviour and suspected multiple liquid or amorphous phases. Several of these are known to form not only close packed, but low-coordinated crystals as well,\cite{fomin_frenkel_pre,fomin_frenkel,fomin_2011,Dzugutov_2014,barbosa_06} moreover, it is possible to design such models using inverse statistical mechanical methods and targeting specific ground state structures.\cite{IP_design_Truskett} However, the thermodynamic stability of these phases, especially at finite temperature, is not always well understood nor established unequivocally. In the current work we demonstrate the importance of exhaustive phase space search in clarifying phase behaviour on the example of the Jagla model.

%In this paper we revisit and contest this claim by performing an in-depth and exhaustive phase space sampling (by using Nested sampling)\cite{EfficientSampling,pt_phase_dias_ns} in combination with a novel method allowing for simulating phases coexistence in dense hard particle systems. Our results allow us to propose an updated phase diagram: we identify a range of previously unexplored phases (some showing surprising similarities to the intertwined structure of the ice VI phase), which raises questions on the very stability of the different liquid phases that the Jagla model is renown for.

%Thus, understanding the properties of the liquid-liquid phase boundary and especially the behaviour around the liquid-liquid critical point is paramount in understanding not just the liquid state of materials, but also the dynamical and thermodynamical anomalies.  

%Only a few of such isotropic models have been suggested in the literature, the soft-core double step potential\cite{soft-core_LL1_2002,soft-core_LL_2004} and the hard-core two-scale ramp model suggested by Jagla.\cite{jagla}
%In the current work we revisit the Jagla model by using the nested sampling technique, known for its unique advantages in studying phase transitions, to calculate properties around the liquid-liquid critical point. 

\begin{figure*}[thb!]
\centering
\includegraphics[width=17cm]{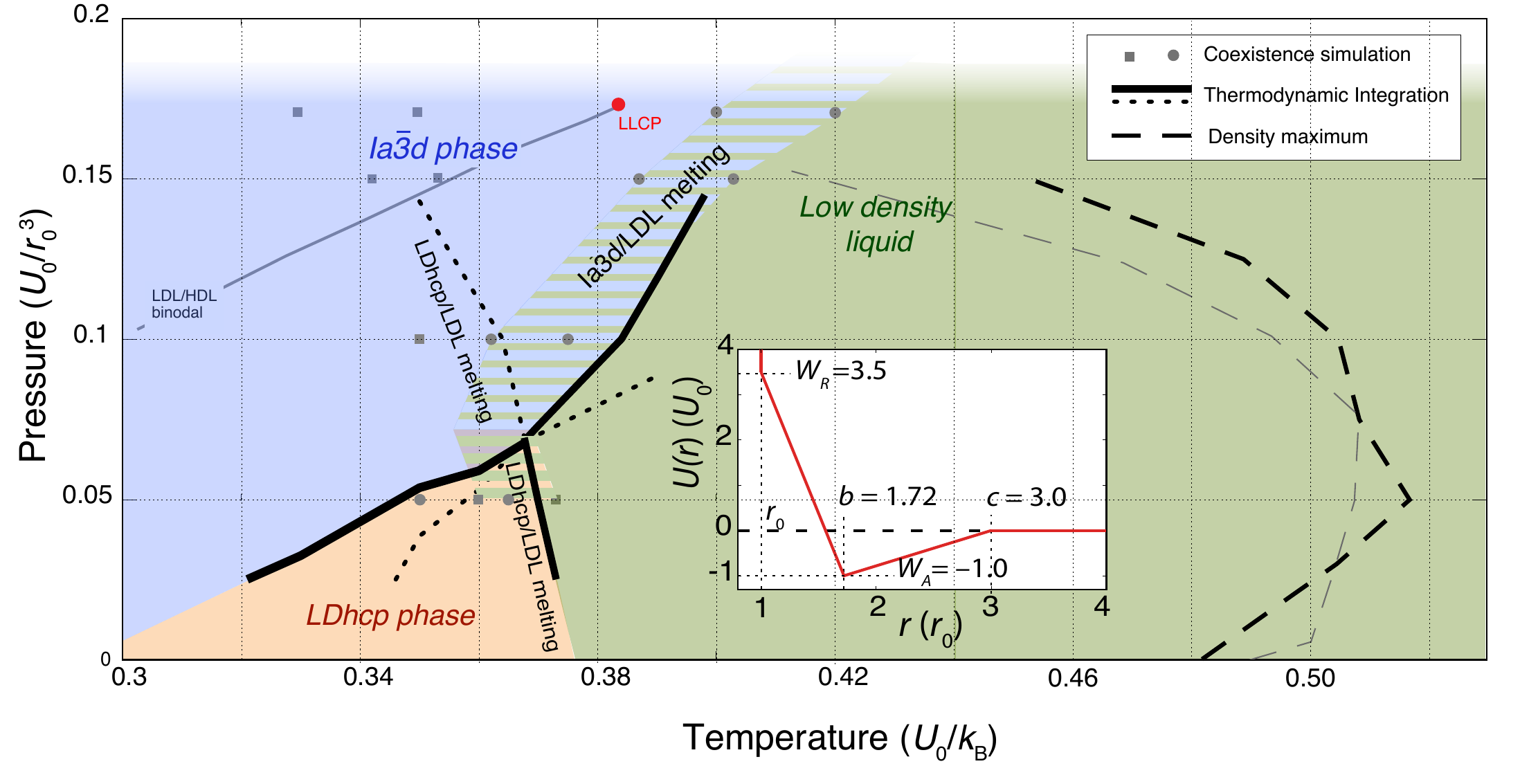}
\caption{$P-T$ Phase diagram of the Jagla model: stability regions of the LDhcp, \iamd{} and LDL phases are highlighted by orange, blue and green shading, respectively. The binodal line between the LDL and HDL phases (taken from Ref.\cite{Ricci}) are shown by a solid grey line, with the LLCP shown by a red circle. Phase transition boundaries calculated by grand canonical coexistence simulations are shown by grey symbols, phase boundaries calculated by thermodynamic integration are shown by black lines, with the dotted sections showing the extension of boundaries in the metastable region. Density maxima of the liquid are shown by black and grey dashed lines from our nested sampling runs and from Ref.\cite{WildingGibson}, respectively. The inset shows the Jagla model with the parametrization used.}
\label{fig:phase_diagram}
\end{figure*}

%%%%%%%%%%%%%%%%%%%%%%%%%%%%%%%%%%%%
% Jagla parametrisation
%%%%%%%%%%%%%%%%%%%%%%%%%%%%%%%%%%%%
The Jagla model is the combination of a hard-sphere core and two linear functions accounting for the repulsive and the attractive ramps, as depicted in the inset of Fig.~\ref{fig:phase_diagram}.
We note that the locations of phase transitions show a strong dependence on the parameterization and the implementation particulars, although the qualitative features of the phase diagram remain unchanged\cite{wildingMagee,Ricci,Xu_glassy,Xu_LL_dyn,Xu_thermo,Luo_slope,WildingGibson,Xu_review,Puertas_jagla_density}.
For details of the Jagla potential used in this work and for the derived Jagla units, we refer the reader to the inset of Fig.~\ref{fig:phase_diagram} and to the Supplementary Material\cite{supp}.
%This parametrisation has been used in several studies, concluding that the model has a liquid-liquid critical point (LLCP) at $P_c = 0.173$ and $T_c = 0.38$, which lies above the melting line.\cite{wildingMagee,Ricci} 
%Studies using the discrete MD method, usually approximating the model with a series of step functions, giving the potential parameter $W_R$ a slightly higher value of 3.5.\cite{Xu_glassy,Xu_LL_dyn,Xu_thermo,Luo_slope} 
%Although it has been shown how changing the position and the ratio of the potential ramps effect the location of the LLCP on the phase diagram\cite{WildingGibson,Luo_slope}, the effect of the stepwise approximation has been largely dismissed, despite it being much more significant.  
%The LLCP of the step-function model is reported to be at a much higher pressures, $P_c = 0.243 \pm 0.003$, $T_c = 0.375 \pm 0.002$.\cite{Xu_review,Luo_slope}. 
%In the current work we use the original linear functions with parameters $W_R=3.5$, $b=1.72r_0$ and $c=3.0r_0$.

In the original work of Jagla\cite{jagla}, both liquid phases were presumed to be thermodynamically stable, because no sign of crystallization was seen in the simulations,
and no further effort had been made to identify possible crystalline structures. 
The heuristic criteria of Hansen-Verlet\cite{Hansen-Verlet} and Lowen\cite{Lowen} were also applied, using the structure factor to conclude that the LLCP is in the stable region. 

In studies where the structure of the solid phase is identified, the hexagonal-close-packed (hcp) crystal,\cite{Xu_thermo,WildingGibson} or at certain potential parameters and at very low pressures, face-centred-cubic (fcc)\cite{WildingGibson,lomba_TI} have been reported.
More exhaustive enumerations of further potential structures of the Jagla model have only been discussed in the two-dimensional system.\cite{jagla_2D_99,2D_Kryuchkov} 

Lomba et al performed detailed calculations on the location of the melting line and the two triple points between fcc solid-LDL-vapour and fcc solid-LDL-HDL.\cite{lomba_TI}
%Interestingly
However, most other studies provide only an estimate to the temperature at which the solid-liquid first-order phase transition occurs, either by giving an upper limit to the melting line by gradually heating a system initially consisting of a crystalline configuration, or estimating conditions where both the liquid and solid phases remain stable within the same simulation cell.\cite{Xu_thermo}
%It is generally agreed however, that the slope of the melting line is negative, and the specific volume of the solid is larger than the volumes of either of the liquid phases.\cite{WildingGibson}

%Where the structure of the solid phase is identified, it is reported as the hexagonal-close-packed (hcp) crystal,\cite{Xu_thermo,WildingGibson} or at certain potential parameters and at very low pressures, face-centred-cubic (fcc).\cite{WildingGibson} 
%Since the potential has no directionality, the assumption that a close packed structure is the most likely crystal arrangement seems viable, and 

%%%%%%%%%%%%%%%%%%%%%%%%%%%%%%%%%%%%
% Nested sampling
%%%%%%%%%%%%%%%%%%%%%%%%%%%%%%%%%%%%

In order to fill this gap in our knowledge of the phase behaviour of the Jagla model, we performed an exhaustive study of its potential energy landscape, employing the nested sampling (NS) method,\cite{Skilling,ConPresNS,EfficientSampling,pt_phase_dias_ns} which allows unbiased exploration without the need of prior knowledge of the stable structures. NS simulations were carried out in the pressure range
between $p=0.003-0.25~U_0/r_0^3$, allowing us to determine thermodynamical phase stability at arbitrary temperatures.
We also performed thermodynamic integration and grand canonical ensemble simulations to accurately determine phase transition conditions\cite{supp}. Our results are summarised in Fig.~\ref{fig:phase_diagram} showing the revised pressure-temperature phase diagram of the Jagla model.
%One of the crucial advantages of this technique is that that no prior knowledge of the stable structures is necessary. 

At pressures below $p=0.05$, our results show excellent agreement with previous findings,\cite{WildingGibson} confirming that the liquid freezes into a hcp solid, which has a lower density than that of the liquid, thus the slope of the melting line is found to be negative. It has been found that the density of the Jagla liquid, similarly to that of liquid water, reaches a maximum when heated isobarically\cite{WildingGibson,lomba_TI}, a behaviour accurately reproduced by our NS simulations.

\begin{figure}
\centering
\includegraphics[width=8cm,angle=0]{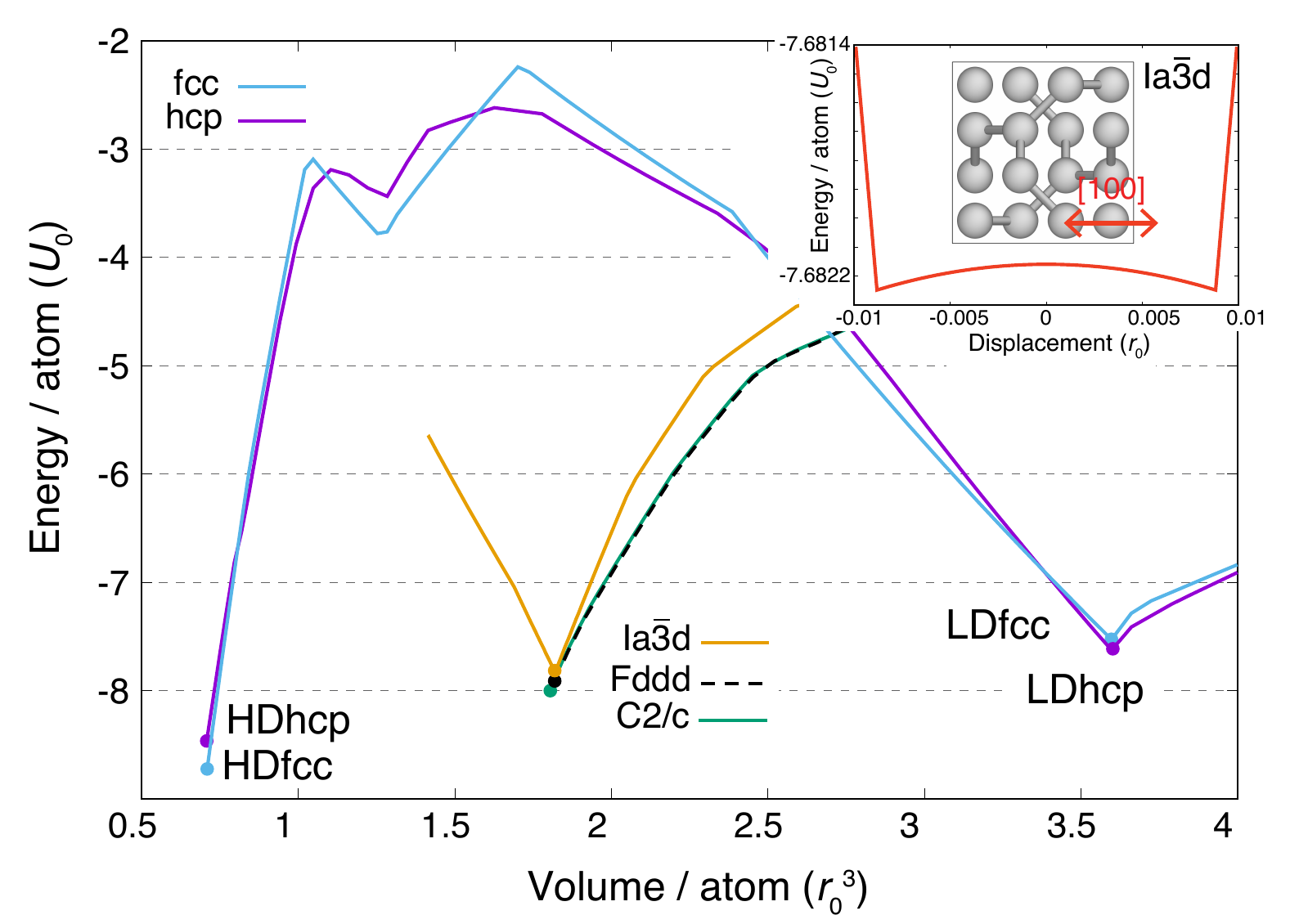}
\caption{Energy dependence on volume, for the perfect crystalline structures. To change the volume, the unit cells were isotropically scaled, all lines shown until the neighbour distances reached the hard-sphere limit. Solid circles indicate the volumes at the energy minimum.
The inset shows the energy of the Jagla \iamd{} crystal as a function of a displacement of an atom along the $[100]$ direction.}
\label{fig:volume_energy}
\end{figure}

Unexpectedly, NS identified a novel, thermodynamically stable, high-pressure solid phase, which has a density at least 30\% higher than the liquid. This structure has a primitive cubic unit cell of 8 atoms, and has an \iamd{} symmetry.  We provide an overview of the energetics of the stable crystalline phases of the Jagla solid in Fig.~\ref{fig:volume_energy}.

We have found that the Jagla \iamd{} crystal is dynamically unstable, as illustrated in the inset Fig.~\ref{fig:volume_energy} for the displacement of an atom in the $[100]$ direction.
As a result, at temperatures above $T=0.22$, the average structure is \iamd{}, but at lower temperatures the structure undergoes two types of distinct symmetric distortions, transforming into lower energy crystalline phases with either $Fddd$ or $C2/c$ symmetries.
The density of the $Fddd$ and $C2/c$ phases is slightly lower than that of the \iamd{}, and upon compression the energy of these phases decreases monotonically until the first neighbour shell reaches the hard sphere limit at the interatomic separation of $r_0$, where no further compression is possible. 

Regarding the low-temperature region of the phase diagram, we found that by isotropically compressing the close-packed hcp and fcc structures, two distinct stability regions can be identified. 
At low densities, the first neighbour-shell is positioned at the distance of the potential minimum, at $b=1.72r_0$, whereas at very high densities the interaction energy is further lowered, reaching the global minimum structure when the interatomic distances at the first neighbour shell are at the hard-sphere limit.
Such iso-structural transitions are known for potential models with multiple characteristic distances.\cite{fomin_review,jagla_2D_99}
We expect that the high-density closed pack phases (HDfcc and HDhcp) are only stable at very low temperatures, as the phase space associated with these highly compressed structures is restricted by the hard-particle repulsive interaction.
Though we saw no sign of the fcc stacking or the extreme high density phases in our calculations, the scaling of the structures suggest that both the HDhcp and HDfcc are much lower in enthalpy, as seen in Figure~\ref{fig:volume_energy}.
The relative stability of hcp and fcc, or potentially other stacking variants, depend on the choice of potential parameters\cite{LiviaPolytypism, Ackland_stacking}, leading to HDfcc being the ground state structure for the parameterization employed in this work.

%Interestingly, while at finite temperature the atoms vibrate around the Ia-3d crystalline sites on average, at very low temperatures the the atoms tend to populate positions away from these sites by $0.01r_0$. 
%The reason for this is that the Ia-3d crystalline sites corresponds to a small potential maxima at low temperature.
%In agreement with this, the nested sampling calculations suggest a solid-solid phase transition at lower temperatures (at around $T=0.22$), where the symmetry of the Ia-3d structure breaks and structure with either Fddd (Wyckoff 70, 8 atoms in the primitive cell) or a C2/c (Wyckoff 15, unit cell of 8 atoms) symmetry is formed. 

Having identified the novel \iamd{} crystalline structure, we carried out thermodynamic integration and coexistence simulations to accurately determine its region of stability.  
Both of these methods suggest that from the pressure $p=0.07$, the \iamd{} structure becomes more stable than the LDhcp phase, with the triple point between the three phases (LDL, LDhcp, \iamd{}) estimated in the temperature region $T=0.36-0.37$.
According to our calculations, above the triple point pressure the melting temperature of the \iamd{} phase increases steeply, and freezing occurs at significantly higher temperatures than the LDL-HDL binodal in the entire pressure range.
We have explicitly compared the Gibbs free energy of the HDL to that of the \iamd{} phase and found that the HDL does not become more stable, even at higher temperatures. Hence the LLCP, in contrary with previous assumptions, is buried deep in the metastable region of the phase diagram.

\begin{figure}[!h]
\centering
\includegraphics[width=\linewidth]{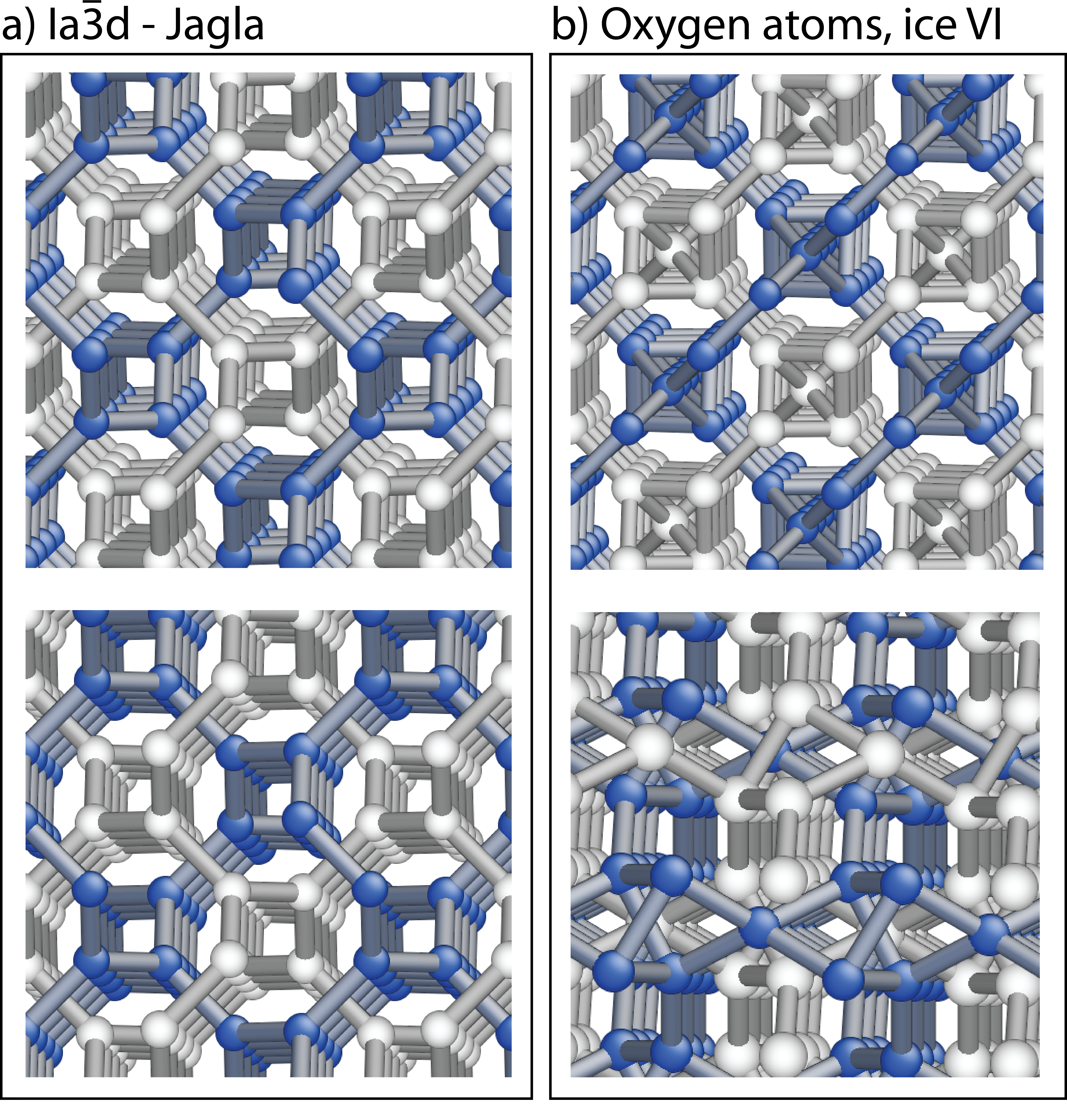}
\caption{Snapshot of the \iamd{} structure of the Jagla model (left) and the oxygen atoms in the ice VI structure (right). Atoms are colored to distinguish the two interpenetrating neighbor networks.
Structures are shown from the $[100]$ (top panel) and $[010]$ (bottom panel) crystallographic directions.}
\label{fig:snapshot}
\end{figure}

\begin{figure}[!t]
\centering
\includegraphics[width=6.3cm,angle=90]{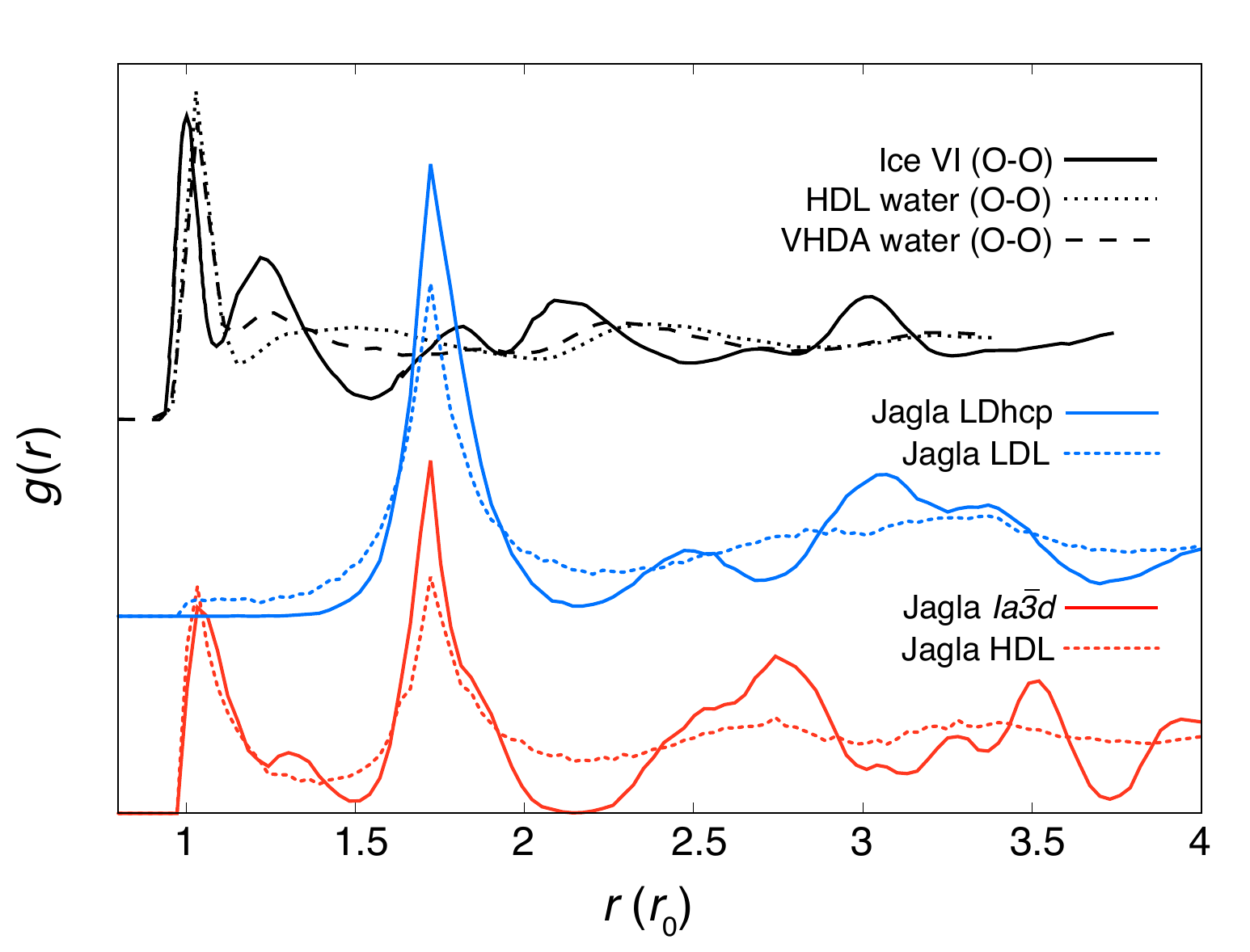}
\caption{RDFs of the LDL/LDhcp phases at ($T=0.36$,$P=0.05$) and the HDL/\iamd{} phases at ($T=0.36$,$P=0.16$). For reference, oxygen-oxygen radial distribution function of the Ice VI phase, high density liquid and the very high density amorphous (VHDA) water are also shown (Refs\cite{tanaka_iceVI,duki_hdlMD}), scaled to align the first peak of the Ice VI phase with that of the Jagla model.}
\label{fig:rdf_jagla}
\end{figure}

The microscopic structure of the \iamd{} phase of the Jagla solid, depicted in the left panels of  Fig.~\ref{fig:snapshot}, is characterised by each atom having three nearest neighbors, which form two interpenetrating networks, each of which is identical to the $K_4$ structure proposed for a range of materials, such as boron\cite{boron_k4}, phosphorus\cite{P_k4} and hydrogenated carbon\cite{CH_k4}.
A strikingly similar structure has been observed in the ice VI phase of water\cite{ice_vi_kamb}, as shown in the right panels Fig.~\ref{fig:snapshot}, where the water molecules also form an intertwined hydrogen bonding network. 
Interestingly, ice VI is one of the two stable crystalline structures which are suspected to hide the HDL and VHDL phases of water on the phase diagram.\cite{water_PD,water_PD_hdl_exp}
We argue that the existence of a stable high density crystalline phase highlights yet another curious qualitative similarity between the Jagla matter and water.
Their radial distribution functions (RDF), after rescaling, are compared in Fig.~\ref{fig:rdf_jagla}.
In the case of the Jagla model, there are two competing local structural features, one characterized by a closed-packed first neighbor shell located around the potential minimum, while the other features a more open structure with three neighbors almost at the hard-core contact limit.
These are observed individually in the corresponding solid phases, LDhcp and \iamd{}, but appear simultaneously in the liquid phases.
This further highlights the analogy of the local structures observed in water and the Jagla matter, and we also note the signature of the intercalated networks appearing as a second peak in the RDF, around $r=1.3$ of Ice VI and the Jagla \iamd{} phases.

The fact that a new, structurally complex phase was found for such a simple model potential that is thermodynamically stable in a large pressure-temperature region demonstrates that relying on just physical intuition may be insufficient in predicting phase diagrams, and without an exhaustive search of the phase space by appropriate computational tools, crucial properties of the system can remain hidden.
Our simulations have identified a previously unexplored crystalline phase, refuting the commonly held presumption that both the high and low density liquid phases of the Jagla matter, and the transition between them are thermodynamically stable.
This raises a series of further questions on whether reparametrization of the model could lower the free energy of the high density liquid and move the LLCP in the stable region of the phase diagram.
This new, hitherto missing piece of the liquid-liquid polyamorphism puzzle provides additional support, and a potential route of generalization of the two-state model\cite{Gallo:2016fd} that has been suggested as a possible explanation for the phenomena.
%, or whether there are unknown crystalline phases yet to be discovered in other promising models of the LL transition.

%These results demonstrate that relying on our physical intuition can be insufficient in predicting phase diagrams, and without an exhaustive search of the phase space by appropriate tools, crucial properties of the system can remain hidden. 
%The unexpected similarities to the ice VI phase suggest...???
%What conclusions can we draw from the similarity to the ice phase? 
%Is there a general rule? Are all the LL phase transitions are based on the same phenomena?

\section*{acknowledgement}
L.B.P. acknowledges support from the EPSRC through an Early Career Fellowship (EP/T000163/1).
This work used the Cirrus UK National Tier-2 HPC Service at EPCC (http://www.cirrus.ac.uk) funded by the University of Edinburgh and EPSRC (EP/P020267/1).
Additional computing facilities were provided by the Scientific Computing Research Technology Platform of the University of Warwick.
Coexistence simulations were performed with the HOOMD-blue package\cite{hoomd_anderson2008,hoomd_glaser2015,hoomd_anderson2016}, atomic structures were visualised using AtomEye\cite{Li:2003wy}. The authors would like to thank James Kermode for stimulating discussions.

\bibliographystyle{apsrev4-2}
\bibliography{jagla_bib}% Produces the bibliography via BibTeX.

\end{document}